\documentclass[useAMS,usenatbib,usegraphicx]{mn2e}
\usepackage{epsfig,psfig}
\usepackage{amsmath,wasysym}
\usepackage{amsfonts}
\usepackage{amssymb}
\usepackage{longtable,booktabs}
\usepackage{multicol}
\usepackage{graphicx}
\usepackage{color}
\usepackage{tabularx}
\usepackage{aas_macros}
\usepackage{comment}
\newcommand{\lxlr}{$P_{1.4\mathrm{GHz}} - L_{X}\,$}

\title[H1821]{How unusual is the cool-core radio halo cluster CL1821+643 ?}
\author[Kale and Parekh]{Ruta Kale $^{1}$\thanks{E-mail:
ruta@ncra.tifr.res.in} and Viral Parekh$^2$\\
$^1$National Centre for Radio Astrophysics, T. I. F. R., Post Bag 3, Ganeshkhind, Pune 411007\\
$^2$Raman Research Institute, C. V. Raman Avenue, Sadashivnagar, Bangalore 560080}

\begin{document}


\pagerange{\pageref{firstpage}--\pageref{lastpage}} \pubyear{2016}

\maketitle

\label{firstpage}

\begin{abstract}
Massive galaxy clusters with cool-cores typically host diffuse radio sources called mini-haloes,
whereas, those with non-cool-cores host radio haloes. 
We attempt to understand the unusual nature of the cool-core galaxy cluster CL1821+643 that 
hosts a Mpc-scale radio halo using new radio observations and morphological analysis of its 
intra-cluster medium. We present the Giant Metrewave Radio Telescope (GMRT) 
610 MHz image of the radio halo. The spectral index, $\alpha$ defined as $S\propto \nu^{-\alpha}$, 
of the radio halo is $1.0\pm0.1$ over the frequency range of 323 - 610 - 1665 MHz. Archival {\it Chandra} X-ray 
data were used to make surface brightness and temperature maps.
The morphological parameters Gini, $M_{20}$ and concentration ($C$) were calculated 
on X-ray surface brightness maps by including and excluding the central 
quasar (H1821+643) in the cluster. 
We find that the cluster CL1821+643, excluding the quasar, is a non-relaxed cluster as seen in the 
morphological parameter planes. It occupies the same region as other merging radio halo clusters in 
the temperature- morphology parameter plane. 
We conclude that this cluster has experienced a non-core-disruptive merger. 

\end{abstract}


\section{Introduction}

Radio haloes, relics and mini-haloes are terms that refer to 
diffuse radio sources in clusters of galaxies that are associated with the 
intra-cluster medium (ICM) as a whole and not with individual galaxies. 
Such sources are direct probes of the relativistic electrons and magnetic fields in the ICM.

Radio haloes are Mpc-sized diffuse radio sources located nearly cospatially with the X-ray emission from the 
ICM. Theoretical models invoking reacceleration of seed electrons through 
a cascade of MHD turbulence introduced in the ICM by merger 
best explain the observed properties of radio haloes \citep[see][for reviews]{don13a,bru14}. 
In addition, relativistic electrons that are secondary products of hadronic interactions 
in the ICM also contribute but are not sufficient \citep[e. g.][]{don10}.

Radio mini-haloes are extended sources  
found surrounding bright central galaxies in cool-core clusters and often
 bounded by sloshing cold-fronts \citep[e. g.][]{maz08,2014ApJ...781....9G}. 
 The origin of mini-haloes has been proposed to be a combination of relativistic 
electrons produced in hadronic interactions and reaccelerated via MHD turbulence 
\citep[e. g.][]{git02,zuh13,2015ApJ...801..146Z}.

The occurrence of radio haloes in merging clusters was noticed since the 
early discoveries of these sources \citep{gio99} and possible scaling relations between the level of 
disturbance and their power were considered \citep{buo01}. Observations of large samples of galaxy clusters 
have strengthened the scenario that radio haloes nearly exclusively occur in massive and merging clusters and 
radio mini-haloes in cool-core clusters \citep{cas10,cas13,2015A&A...579A..92K}. 

The recent discovery of a radio halo in the cool-core cluster CL1821+643 challenges the current understanding 
\citep[][hereafter, B14]{bon14}. 

\subsection{The cluster CL1821+643}\label{intro1}
CL1821+643 is a massive, $M_{500}$ = 6.311 $\times$ 10$^{14}$ $M_{\odot}$  \citep{2014A&A...571A..29P}, 
luminous and rich cool core cluster at a redshift of 0.299 \citep{sch92}. 
This cluster was discovered surrounding a highly luminous broad-line  
quasar, namely H1821+643 \citep{sch92}. An FRI type radio source of extent $\sim200$ 
kpc has been found to 
be associated with the central quasar with possibly precessing jets \citep{blu01}. 
\citet{rus10} performed 
detailed modeling of this quasar using {\it Chandra} data and were able to 
subtract the quasar to estimate the internal properties of the cluster itself. 
They calculated a central radiative cooling time of $\sim$ 1 
Gyr which is comparable to other cool core clusters. They also 
measured a radial temperature decline from 9 keV outside of 
the central $\sim$ 200 kpc to about 1.3 keV in the core 
region of 20 kpc. 
In their X-ray imaging and corresponding temperature map, they noticed 
elongated morphology of the cluster in the north-west to south-east 
direction, a few extended arms of emission around the core, an outer swirl 
with cool gas motion, and a possible cold front generated by the 
cool-core sloshing. 
\citet[][]{rus10} conclude that the large-scale 
cluster ICM properties are not substantially affected by the central quasar 
activity; outside of 8 arcsec ($\sim$ 35 kpc), the cluster 
emission is dominant, and the emission from the quasar is negligible. 
Recently \cite{2014MNRAS.442.2809W} analysed thermodynamic properties 
of the cluster CL1821+643 and found that the entropy around the central 
quasar between 30 - 80 kpc is substantially lower than that of other comparable 
massive, strong cool core clusters. 

A radio halo of size $\sim1.1$ Mpc has been discovered in this cluster 
 at 323 MHz with the GMRT and at 1665 MHz with the VLA 
D-array by B14. They constrain the spectral index\footnote{The spectral index, $\alpha$ is defined as $S_\nu \propto \nu^{-\alpha}$, where 
$S_\nu$ is the flux density at the frequency $\nu$.} of the radio halo to the range  
1.04 - 1.1 over frequencies 323 and 1665 MHz. 
They used the morphological parameters, Concentration parameter ($c$), centroid shift ($w$) 
and power ratios ($P_3/P_0$) to characterise its dynamical state based on 
the X-ray map with archival Chandra data. The cluster is found consistent with 
relaxed clusters in the $c - P_3/P_0$ plane but is an outlier in the 
$c - w$ plane due to high $w$. 

In this work we use our GMRT 610 MHz and archival 150 MHz observations to 
study the radio sources in the cluster. We also perform a morphological 
parameter analysis of the cluster using a different set of parameters 
and compare the cluster with a large sample of clusters 
to understand the peculiarities in this cluster that led it to host a radio halo 
while still having properties of a cool-core cluster.

We have adopted $\Lambda$ CDM cosmology with $H_0 = 70$ km s$^{-1}$ Mpc$^{-1}$, $\Omega_\Lambda=0.73$ and 
$\Omega_m = 0.27$. At the redshift 0.299 of the cluster CL1821+643, one arc second corresponds 
to 4.47 kpc. 

\section{GMRT observations and data reduction}\label{rdata}

The cluster CL1821+643 was observed with the GMRT 
at 610 and 235 MHz using the dual frequency mode of 
the GMRT software backend that allows to record one polarization at each frequency band. 
The data at 235 MHz had to be discarded due to  
unstable phases on the phase calibrator and high radio frequency 
interference. 
The 610 MHz data were recorded with a bandwidth of 32 MHz spread over 512 channels 
for a duration of 200 min on target source.
We used the flagging and calibration pipeline `FLAGCAL' \citep{chen13}
for data reduction. This code uses the statistics of median and median absolute deviation for 
flagging. The flux, bandpass and phase calibrator data are flagged and calibrated first. These are used to calibrate 
the target source data. The calibrated target source data are flagged and 
recalibrated if needed. The resulting output FITS file contains flagged and calibrated visibilities on 
all the sources.
These were loaded in NRAO Astronomical Image Processing System (AIPS) for further analysis. 
The data were first inspected to verify the calibration and flagging. Additional minor flagging was carried out 
manually.
The target source data were then split and averaged in frequency appropriately to 
reduce the data volume and at the same time avoid being affected by bandwidth smearing. 
These were then imaged using the AIPS task `IMAGR' and followed by self-calibration.

Images were produced using the final visibilities with a variety of weighting 
schemes for the visibilities. The high and low resolution  (HR and LR, respectively) 
images presented in this paper were obtained as follows.
The high resolution (HR) image of the point sources in the field 
was made excluding the visibilities in the inner 2$k\lambda$ of the uv-plane and 
robust$=0$. 
The clean components of the point sources were subtracted from the visibilities using the task `UVSUB'. 
The resulting visibilities were used to make a low resolution image of the diffuse emission with 
robust$=3.0$ and uv-distances up to 15$k\lambda$. The resulting image was  
was convolved to a circular beam of $30''$ and is referred to as the low resolution  (LR) image.

The amplitude scale is set according to Baars et al. (1977) within the pipeline FLAGCAL used for 
data analysis. This is converted to the Scaife-Heald 2012 scale using a factor\footnote{The scaling factor of 1.046 given for 750 MHz in Table 7 in \citet{1977A&A....61...99B} between KPW \citep{1969ApJ...157....1K} and Baars 1977 scale 
is assumed for 610 MHz. \citet{2012MNRAS.423L..30S} use the RCB scale \citep{1973AJ.....78.1030R} for frequencies above 325 MHz which they state is consistent with KPW.}  of 1.046
\citep[][]{2012MNRAS.423L..30S}. 
An amplitude error ($\sigma_{amp}$) of $10\%$ is assumed at 610 MHz.
The error on the flux density is calculated according to $\Delta S = [(\sigma_{amp}S)^2 + (\sigma_{rms}\sqrt{n_{beams}})^2)]^{1/2}$, where $S$ is the flux density, $\sigma_{rms}$ 
is the image rms noise and $n_{beams}$ is the number of beams in the extent of the source.

\subsection{Radio Images}

The high and low resolution GMRT 610 MHz radio images of the cluster CL1821+643 are presented in 
Fig.~\ref{h1821}. The rms noise in HR and LR images are 0.06 and 0.24 mJy beam$^{-1}$, respectively.
The discrete sources are labelled A and B following the labels 
used by B14. 
Source A can be identified with a galaxy SDSS J182154.96+642117.1 and has been 
classified as a cluster member \citep{sch92}.
Source B is an FR I radio source associated with the central quasar H1821+643 
\citep{blu01}.

\section{X-ray data analysis and images}\label{xdata}
We used {\it Chandra} archival 
data to image the cluster CL1821+643. A summary of the data is given in 
Table~\ref{CL1821+643_X-ray_info}. 
We processed the data with 
CIAO 4.6 and CALDB 4.6.1.1. The \verb"chandra_repro" task was used to 
reprocess all ACIS imaging data. Any high background flares were removed 
 with the task \verb"lc_sigma_clip" (3$\sigma$ clipping). The read out 
strip was removed using the \verb"destreak" task, 
followed by \verb"merge_all" script to combine all data sets. 
The total exposure time of these combined data is 
74.5 ks. All of our event files included the 0.3--7 keV broad energy band 
and 2$''$ pixels binning. We removed point sources around the cluster 
(except the central bright quasar). We divided counts image with exposure map 
 to generate the flux image. Finally we smoothed this image with $\sigma$ = 10$''$ to 
remove zero counts. The final images were used to calculate the morphological 
parameters that serve as proxies to the dynamical state of the cluster.
\subsection{Density and temperature maps}
Temperature map of the ICM is also a good indicator of 
the dynamical activity in the cluster.
We used XMC (X-ray Monte Carlo) technique to generate it 
\cite[see][for details of the method]
{2007ApJ...655..109P, 2007ApJ...670.1010A, 2009ApJ...696.1029A}. 
In this method the spectral and spatial models are used together. 
We used the warm-absorbed APEC (spectral) model along with 
the {\it Chandra} detector (spatial)
 model derived for XMC. 
In this analysis, we kept $n_{H}$ = 0.0403 $\times$ 10$^{22}$ cm$^{-2}$
 \citep{1990ARA&A..28..215D}, $Z_{\odot}$ = 0.3 and $z$ = 0.299 fixed, 
while allowed temperature (T) to vary between $\sim$ 1 to 15 keV. 
In this procedure, given (free) parameters (temperature, spatial coordinates of 
detector, APEC normalisation factor, etc.) of the ICM are iterated using
 the MCMC technique. The final results have a  number of statistical 
samples of acceptable or converged fits which fall into the ``confidence 
region'' of the ICM's input parameters. These well ``fitted'' parameters 
describe the properties of the ICM. In this work, all the results 
derived from the model samples are from the iterations from 500 to 3000, 
where the value of $\chi^{2}$ is reduced (to $\sim$ 1) and stable, and 
is considered to be a converged chain where all corresponding parameters 
have the best-fitting value.  We did not remove quasar from the events file 
for this analysis. Temperature map shows high value because in the XMC method 
multi-temperature plasma overlaps on each other 
 which mimics the `real' situation in ICM plasma \citep{2009ApJ...696.1029A}.

\begin{table}
\centering
\caption{CL1821+643 $Chandra ACIS-S$ X-ray data.}
\begin{tabular}{@{}cccccccc@{}}
\hline\noalign{\smallskip}
ObsID, PI & Date of Observation & Exposure time (ks) \\
\hline\noalign{\smallskip}
9398, A. Fabian & 2008-04-21  &34.2\\
9845, A. Fabian & 2008-04-14 & 24.5\\
9846, A. Fabian & 2008-04-20 &18.3\\
\hline 
\end{tabular}
\label{CL1821+643_X-ray_info} 
\end{table}

\begin{table}
\begin{center}
\caption[]{\label{srctab}Radio sources in the CL1821+643 region.}
\begin{tabular}{ccccccccccccc}
\hline\noalign{\smallskip}
Source & $RA_{J2000}$ & $DEC_{J2000}$ & $S_{610\mathrm{MHz}}$ (mJy)\\
\hline\noalign{\smallskip}
A  & 18 21 54.84 &+64 21 17.00   & $ 110 \pm 11$ \\ 
B  & 18 21 56.99 &+64 20 34.99   & $ 105 \pm 10$ \\
Radio Halo & 18 21 56.99 &+64 20 34.99 & $35.6\pm4.0$\\
\hline\noalign{\smallskip}
\end{tabular}
\end{center}
\end{table}

\begin{figure*}
\centering
\includegraphics[height =8.6cm]{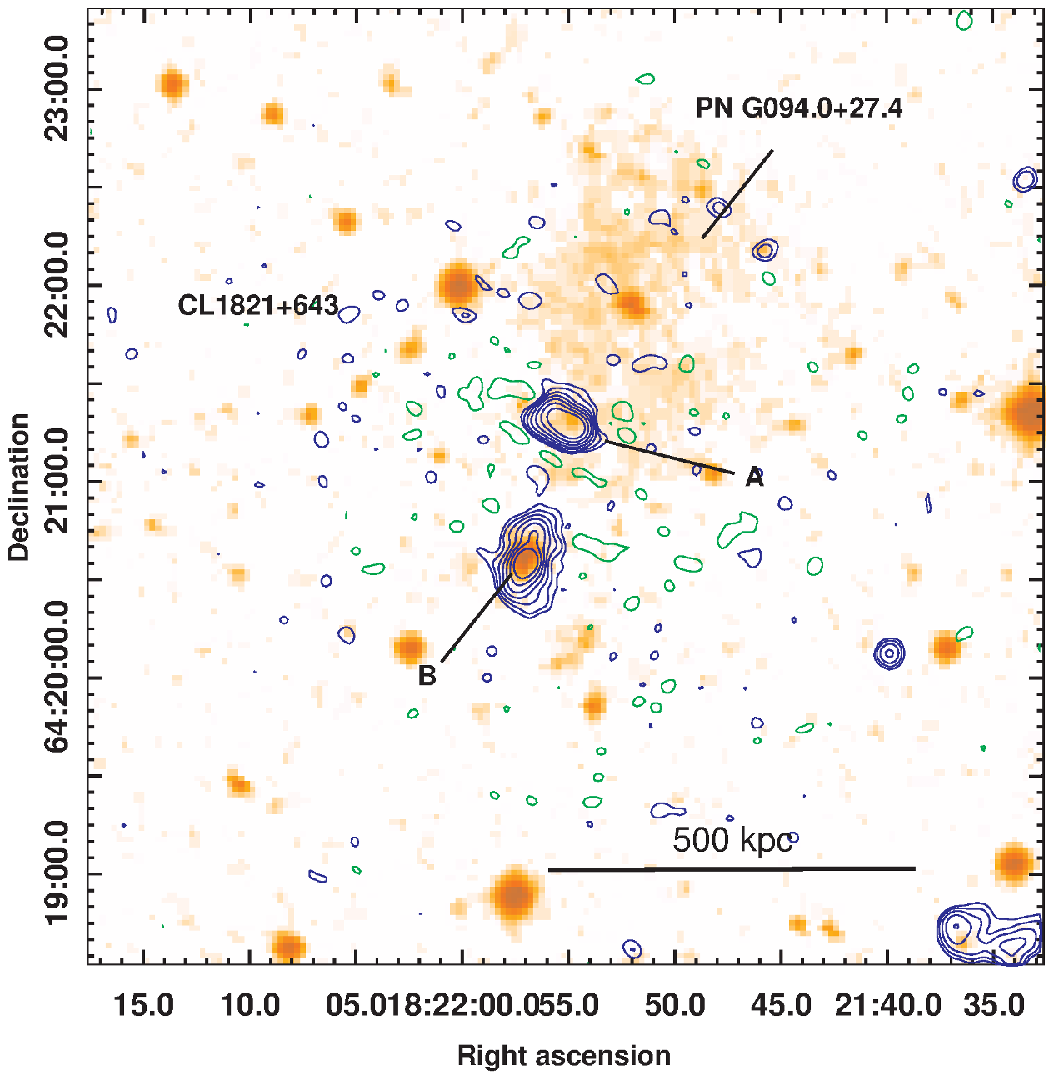}
\hspace{0.2cm}
\includegraphics[height =8.6cm]{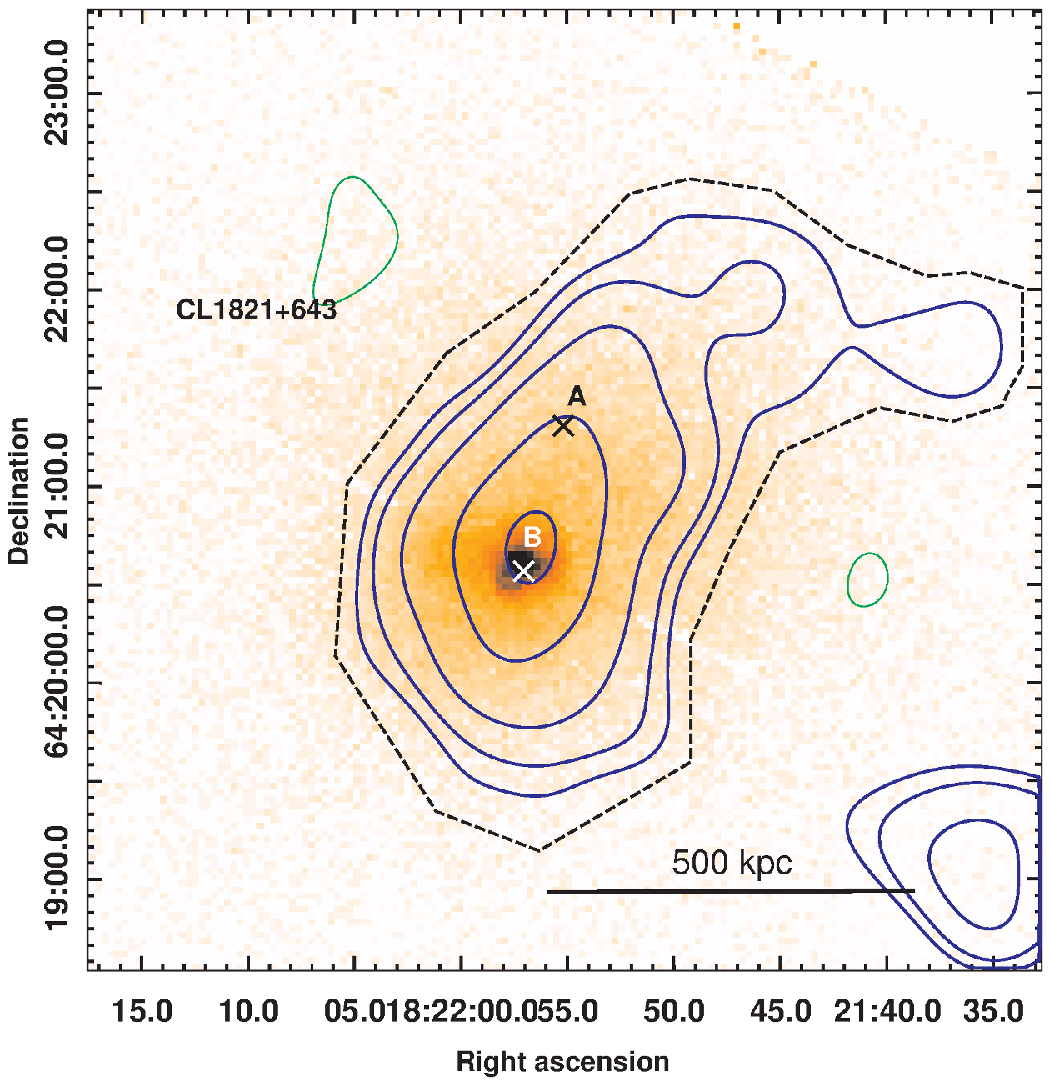}
 \caption{{\it Left--}GMRT 610 MHz HR image in contours is overlaid on the 
DSS R-band optical image. The contours are at 0.24$\times [-1, 1, 2, 4, ...]$ mJy beam$^{-1}$ 
and the synthesized beam is $5.1''\times5.0''$, p. a. -20.2$^{\circ}$ .
  The diffuse emission seen in optical, from the 
planetary nebula is labelled. {\it Right--}{ GMRT 610 MHz LR image shown in contours 
is overlaid on the Chandra X-ray image in colour. The contour levels are $\pm$ 0.9, 1.2, 1.8, 3.0, 4.4 
mJy beam$^{-1}$ and the beam is $30''\times30''$. The flux density of the radio halo was 
measured within the region marked by black dashed line. Crosses mark the positions of sources A and B. 
 In both the panels, 
dark (blue) are +ve and light (green) are -ve contour levels.
A colour version of this figure is available online.}
\label{h1821}}
\end{figure*}

\section{Radio halo in CL1821+643}
\begin{figure}
\includegraphics[width=8cm]{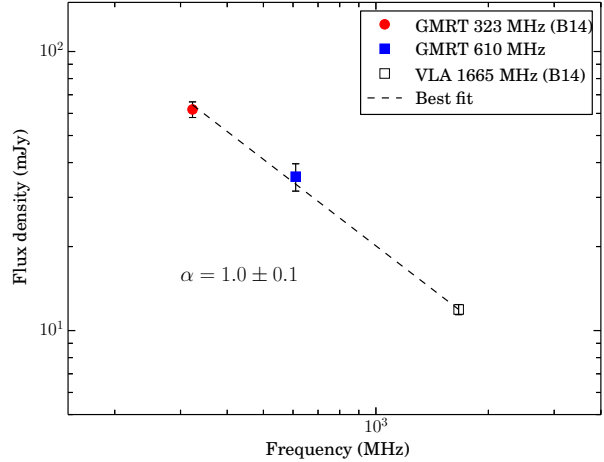}
 \caption{The integrated spectrum of the radio halo in CL1821+643.
 \label{haloespec}}
\end{figure}

The radio halo is detected at 610 MHz and has an extent of $\sim 890$ 
kpc in the north-south and $\sim 450$ kpc in the east-west direction 
(Fig.~\ref{h1821}, right). 
The optical image of the cluster field shows the presence of a 
planetary nebula (PNe) to the northwest of the cluster named, PN G094.0+27.4 
 associated with the white dwarf WD1821+643 (Fig.~\ref{h1821}, left). 
PNe have spectra of thermal nature 
that grow weaker at low frequencies with typical strength of a mJy at 1 GHz \citep[e. g.][]{gur97}. 
No synchrotron emission is known from them though a weak component has been 
discussed in a few cases \citep{casa07}. Therefore we consider that any contamination 
from the PNe near the field of this cluster is below our detection limits.
{The flux density of the radio halo at 610 MHz measured from the low resolution 
image ($30''\times30''$) over the region marked in Fig.~\ref{h1821} (right) 
is $35.6\pm4.0$ mJy. The spectral index ($\alpha$) of the radio halo over the frequency 
range 323 -1665 MHz is $1.0\pm0.1$ 
(Fig.~\ref{haloespec}). The k-corrected radio power of the radio halo is 
$(1.0\pm0.1)\times10^{25}$ W Hz$^{-1}$ according to 
$P_{610\mathrm{MHz}}=4 \pi d_L^2 S_{610\mathrm{MHz}}(1+z)^{(\alpha-1)}$, 
where $d_L$ is the luminosity distance.}
 
Magnetic fields in galaxy clusters are ubiquitous and have been observed to be 
of the order of $\sim0.1 - 1 \mu$G \citep[see][for reviews]{car02,fer08}.  
It is difficult to measure magnetic fields in clusters directly but can be estimated under assumptions. 
Assuming equipartition 
condition, we calculated the magnetic field in the cluster CL1821+643 from the observed 
radio halo emission. 
The minimum energy density $u_{min}$ is given by,
\begin{eqnarray}
 u_{min}\Bigl[\frac{\mathrm{erg}}{\mathrm{cm}^3}\Bigr] = \xi(\alpha,\nu_1,\nu_2) (1+k)^{4/7} 
 (\nu_{0[\mathrm{MHz}]})^{4\alpha/7} \times \nonumber \\ (1+z)^{(12+4\alpha)/7}
  \bigl(I_{0[\frac{\mathrm{mJy}}{\mathrm{arcsec}^2}]}\bigr)^{4/7} (d_{[\mathrm{kpc}]})^{-4/7}
\end{eqnarray}
where $k$ is the ratio of energy in relativistic protons to that in electrons, $\alpha$ is the synchrotron spectral 
index, $\nu_0$ is the frequency at which the surface brightness, $I_{0}$ is measured, $d$ is the 
depth of the source and $\xi(\alpha,\nu_1,\nu_2)$ is a parameter that is a function of the spectral index 
and the lower and higher limits in frequency, $\nu_1$ and $\nu_2$ \citep{gov04}. 
The $K-$correction is included and a filling factor of 1 is assumed in the above equation.
The magnetic field is then given by, 
\begin{equation}\label{beq1}
 B_{eq[\mathrm{G}]} = \Bigl(\frac{24 \pi}{7}u_{min}\Bigr)^{1/2}.
\end{equation}
The depth of the radio halo was assumed to be the mean of the maximum and minimum extents 
($890$ kpc $\times 450$ kpc). {The extent in the plane of sky over which the 
total flux density} was measured encloses 22 beams of $30''\times30''$ each and thus the surface brightness was 
calculated to be, 35.6 mJy / ($22\times$($30\times30$ arcsec$^2$)). For the $\alpha=1.0$, $\nu_1=10$ MHz and 
$\nu_2=100$GHz, $\xi=5.39\times10^{-13}$. The implied magnetic field under these assumptions is 0.63 $\mu$G.
The modified equipartition magnetic field ($B'_{eq}$) based on a limit on the 
minimum Lorentz factor, $\gamma =100$ rather than on the frequency \citep[e.g.][]{bec05}, given by, 
\begin{equation}
 B'_{eq[\mathrm{G}]} \sim 1.1 \gamma^{\frac{1-2\alpha}{3+\alpha}}_{\mathrm{min}} B_{eq}^{\frac{7}{2(3+\alpha)}},
\end{equation}
where $B_{eq[\mathrm{G}]}$ is from eq.~\ref{beq1}, is 1.3 $\mu$G for the radio halo.

\section{Morphological parameters}\label{morfparms}
The X-ray emitting gas in galaxy clusters carries signatures of dynamical activity that  
manifests itself as distortions in the X-ray surface brightness images. 
We computed three morphology parameters, Gini, $M_{20}$ and Concentration ($C$) 
 using the {\it Chandra} X-ray image of CL1821+643 to characterise the degree of disturbances 
in its ICM. These three parameters are known to be effective in seggregating 
galaxy clusters according to the level of disturbances in them 
\citep[][hereafter, P15]{2015A&A...575A.127P}. 
Gini is a measure of the flux distribution among the image pixels; its value is 0 
if the flux is equally distributed among the pixels and is 1 if most of the flux is 
contained only in a small number of pixels \citep{2004AJ....128..163L}. The 
moment of light, $M_{20}$, is the normalised second order moment of the relative 
contribution of the brightest $20\%$ of the pixels \citep{2004AJ....128..163L} 
and is a measure of the 
spatial distribution of bright cores and substructures in the cluster. 
The typical values of $M_{20}$ are  
between -2.5 (very relaxed) to -0.7 (very disturbed) (see P15).
The parameter $C$ is a measure of concentration of the flux in the cluster that depends on the 
ratio of the radii at which $80\%$ and $20\%$ of the cluster flux is found 
\citep{2003ApJS..147....1C}
and has a minimum value of 0.0. We point the reader to P15 and the references 
therein for the details 
of the parameter calculations.

The parameters for CL1821+643 were calculated with and without the central quasar in order 
to understand the 
effect of the quasar on the morphology.
 For the subtraction of the quasar, we excluded 25 kpc central region around RA 18h21m57.239s  
DEC +64d20m36.22s similar to that in B14. 
A flux weighted centre was found in both the cases using 
an iterative method described in P15. The centres with and without 
the quasar were found to be separated by $2.83''$ which is $13$ kpc at the cluster.
A region of 500 kpc around the centre was used for 
the calculation of the morphological parameters.
In Table~\ref{CL1821+643_X-ray_morph} the parameter 
values along with 1$\sigma$ uncertainty are stated.

\begin{table*}
\centering
\caption{Morphology parameter values with and without core. The change in the parameter values (percentage) 
are also given for each parameter in adjacent parentheses.}
\begin{tabular}{cclllccc}
\hline \noalign{\smallskip}
Cluster&Method&Gini ($\% $change) &$M_{20}$ ($\% $change)& $C$ ($\% $change) \\
\hline\noalign{\smallskip}
CL1821+643&With quasar & 0.66$\pm$0.011 & -2.17$\pm$0.48 & 1.72$\pm$0.51\\
CL1821+643&Without quasar (25 kpc) & 0.58$\pm$0.005 (12) & -1.85$\pm$0.33 (15) & 1.38$\pm$0.41 (20)\\
\hline
A2597 & &  0.78 $\pm$ 0.0027 & -2.42 $\pm$0.25 & 2.23 $\pm$0.25  \\
A2597 & 25 kpc core removed&  0.75 $\pm$ 0.0025 (4)& -2.10 $\pm$0.20 (13) & 2.03 $\pm$ 0.27 (9)\\
RXJ1504& & 0.80 $\pm$ 0.0050 & -2.47 $\pm$ 0.45 & 2.08 $\pm$ 0.35\\
RXJ1504 & 25 kpc core removed & 0.76 $\pm$ 0.0038  (5)& -2.0  $\pm$ 0.28 (19) & 1.81 $\pm$ 0.32 (13)\\
\hline 
\end{tabular}
\label{CL1821+643_X-ray_morph} 
\end{table*}

\section{Discussion}\label{disc}
The cluster CL1821+643 presents an intriguing case of a radio halo in a cool-core cluster. 
The properties of the radio halo and the morphological parameters of the cluster are 
compared with other galaxy clusters with and without radio haloes and those with mini-haloes 
in order to understand the uniqueness of this cluster.

\begin{figure}
\centering
\includegraphics[scale=0.450]{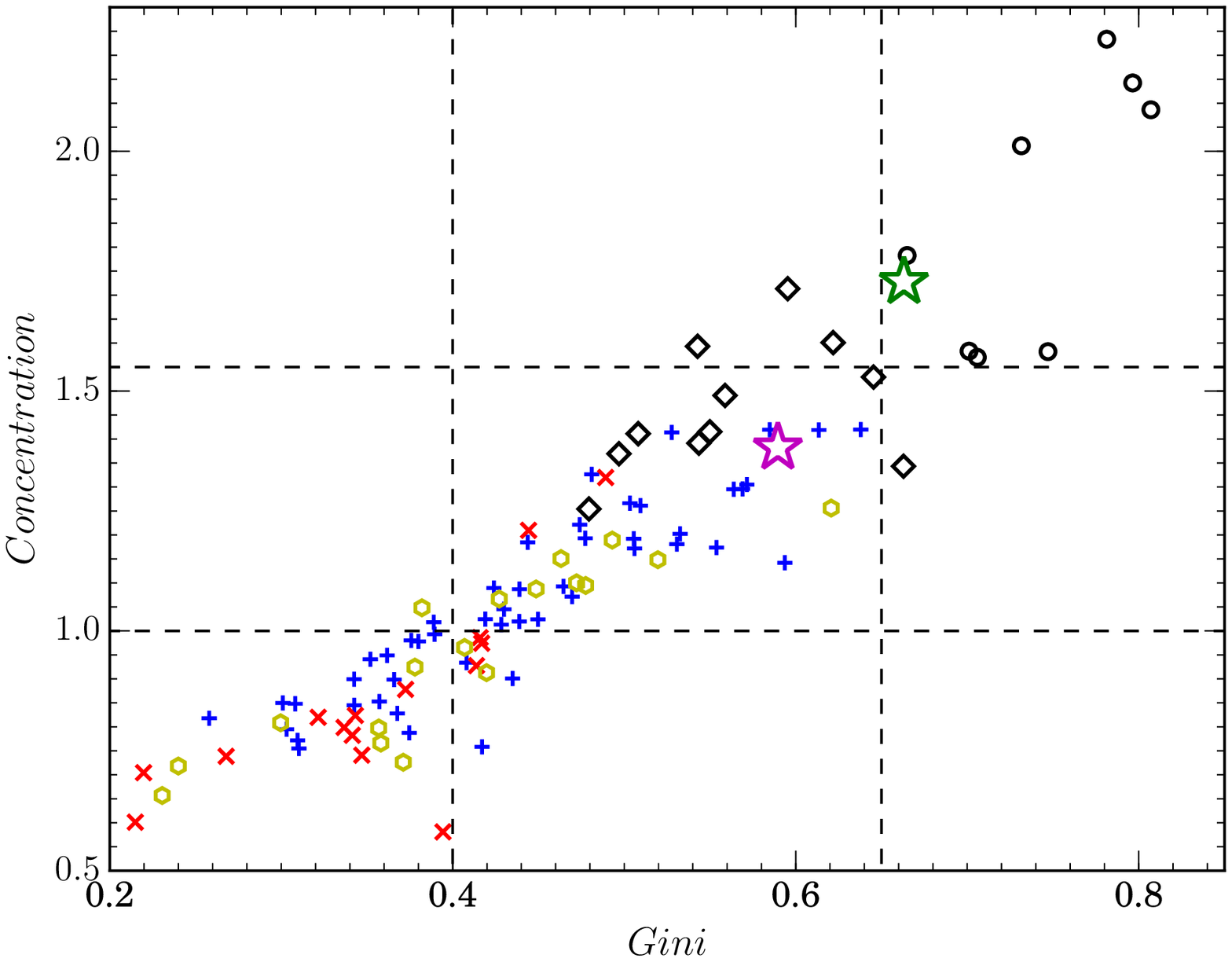}
\includegraphics[scale=0.450]{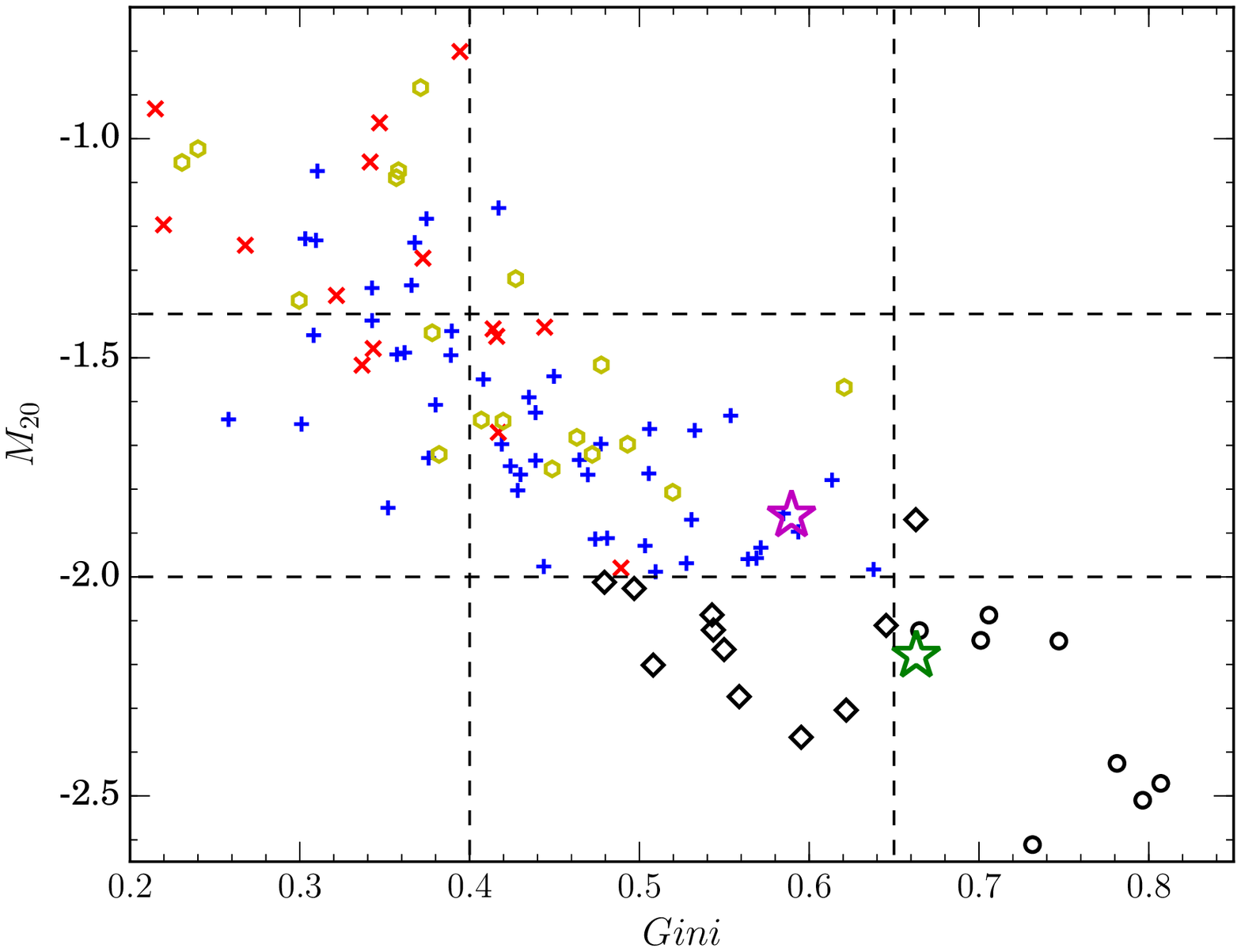}
\includegraphics[scale=0.450]{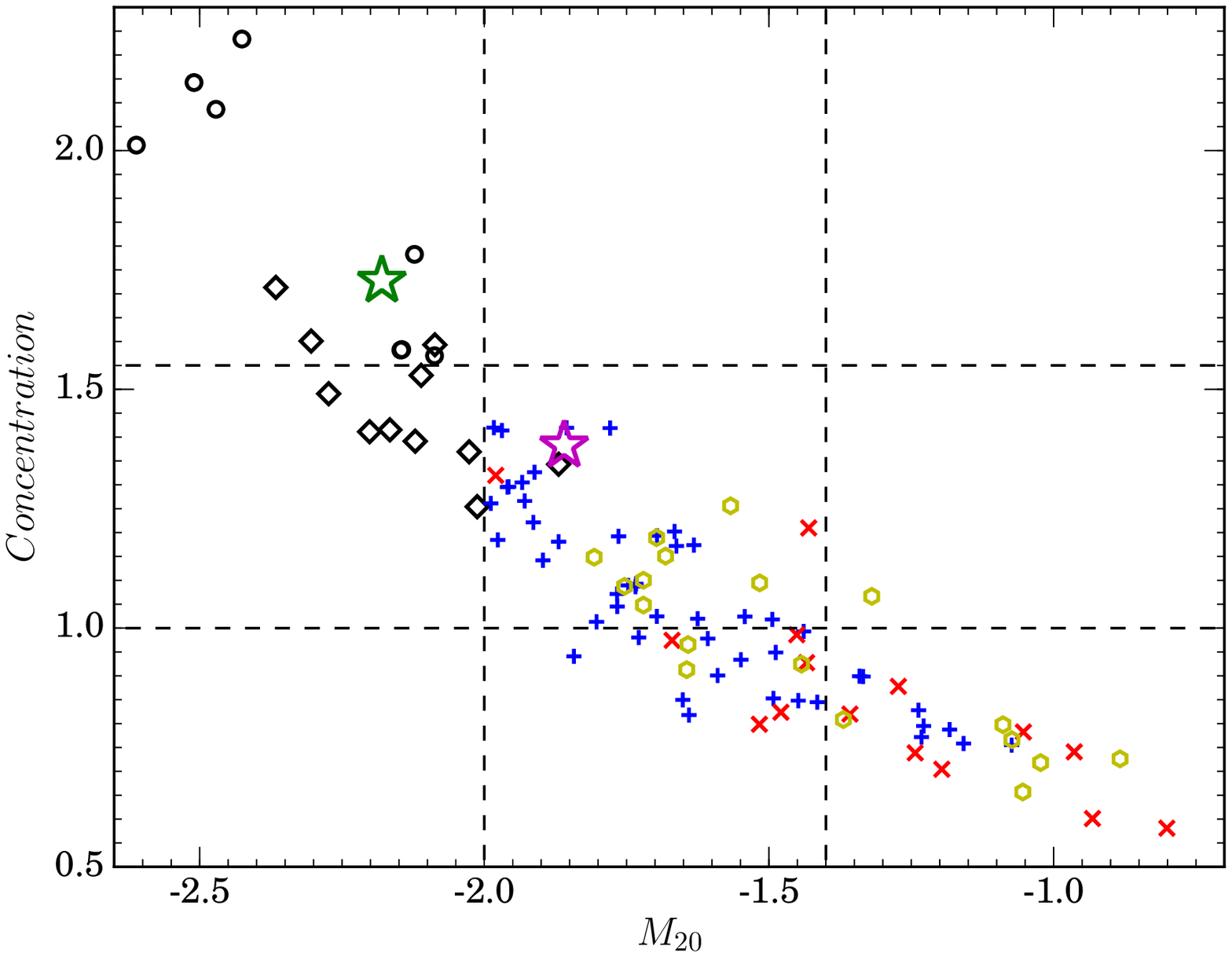}
\caption {\small \small The cluster CL1821+643 is shown in morphology 
parameter-parameter planes along with the P15 sample of clusters. The green 
and the magenta stars denote CL1821+643 with and without the central quasar, respectively.
The remaining sample clusters are shown according to their classification --
 circles are
`strong relaxed', diamonds are `relaxed', plus signs are `non-relaxed' and 
crosses are `strong non-relaxed' clusters. The yellow hexagons are clusters that 
host radio haloes ($z > 0.1$) taken from \citep{gio09} and known to be merging clusters.
}
\label{CL1821+643_Xray1}
\end{figure}

\subsection{Evidence for dynamical activity in CL1821+643}
The morphological parameters of galaxy clusters quantify the disturbances in the 
ICM of the cluster. We compare the parameters calculated for CL1821+643 with 
a sample of 85 clusters with deep Chandra observations 
 for which morphological parameters are available in P15. 
Among the sample of 85 clusters, there are 49 low-$z$ (0.2 -- 0.3) and 36 high-$z$ (0.3 -- 0.8) clusters  
with X-ray flux, $f > 1.4\times10^{-13}$ erg s$^{-1}$ cm$^{-2}$ and $L_{X}> 3 
\times 10^{43}$ erg s$^{-1}$. 

P15 have classified the clusters based on the morphological parameters into 
four categories of their dynamical states: \\
{\bf Strong Relaxed (SR)} \\
$C > 1.55$, $M_{20}<-2$ and Gini$>0.65$;\\
{\bf Relaxed (R)} \\
$1.0 < C < 1.55$, $-2.0 < M_{20}< -1.4$ and $0.40<$Gini$<0.65$;\\
{\bf Non-Relaxed (NR)}\\
$C < 1.0$, $M_{20}>-1.4$ and  $0.40<$Gini$<0.65$ and \\
{\bf Strong Non-Relaxed (SNR)}\\
$C < 1.0$, $M_{20}>-1.4$ and Gini$<0.40$. 

In Fig.~\ref{CL1821+643_Xray1}, the cluster CL1821+643 is shown along with the P15 sample 
of clusters in the morphological parameter planes.

The SR clusters are well separated from SNR clusters.
CL1821+643 with the central quasar falls in the SR cluster category 
while without the quasar, falls in the NR cluster category. In the first 
case, all three parameters indicate that CL1821+643 is a relaxed cluster, while 
in the second case, all three parameters indicate that CL1821+643 is  
nearly a non-relaxed cluster. This indicates that 
there are disturbances or substructure present in CL1821+643 
outside of the central region, possibly from a high impact parameter merger
that kept the central cool-core unaffected.

In the morphological parameter analysis of B14, the parameter, centroid shift 
showed evidence of dynamical activity and the concentration (a different definition than 
used in this work) and power ratios implied an undisturbed cluster. The morphological 
parameters used here show further evidence for the presence of dynamical activity 
based on the morphology in X-rays. The distribution of galaxies with photometric redshifts between 
0.25 - 0.35 selected from the SDSS shows an assymetric distribution around the central 
quasar \citep{2011ApJ...737...64A}. This supports the disturbance in the cluster 
as indicated by the morphological parameters but needs deeper optical studies with 
spectroscopic identification of cluster members and optical substructure analysis.

\subsubsection{Effect of removing the central quasar}
The cluster CL1821+643 can be classified as a merging cluster based on the morphological parameters 
calculated after subtraction of the central quasar. However it is necessary to test if other similar 
clusters containing bright central sources also turn out to be merging if the central source is removed or 
they continue to be classified as cool-cores.
 We selected two strong relaxed clusters, A2597 and RXJ1504 (RXCJ1504-0248) from P15 sample and re-calculated 
the parameters 
for these two clusters by removing central 25 kpc region around their peaks in X-rays. 
These two clusters were chosen for comparison as they are identified as strong relaxed clusters 
among the P15 sample clusters and their parameter values are similar to the CL1821+643 with the quasar.
The parameters and the differences in each are given in Table ~\ref{CL1821+643_X-ray_morph}. 
The change in the parameters when calculated after removing the core is between 12 to 20$\%$ in the case 
of CL1821+643 as compared to 4 to 13$\%$ for A2597 and 5 to 19$\%$ for RXCJ1504.
This excercise illustrates that in CL1821+643 all the three parameters indicate morphological disturbance 
after removing the central 25 kpc region affected by the quasar. The other two 
strong cool-core clusters do not show a consistent deviation of morphological parameters towards 
disturbance after the core is removed. Thus CL1821+643 is different from other cool-core clusters 
with bright central X-ray sources.

\subsubsection{Temperature map and radio halo morphology}\label{rhmorph}
\begin{figure}
\includegraphics[trim=0 20 0 80,width=\columnwidth]{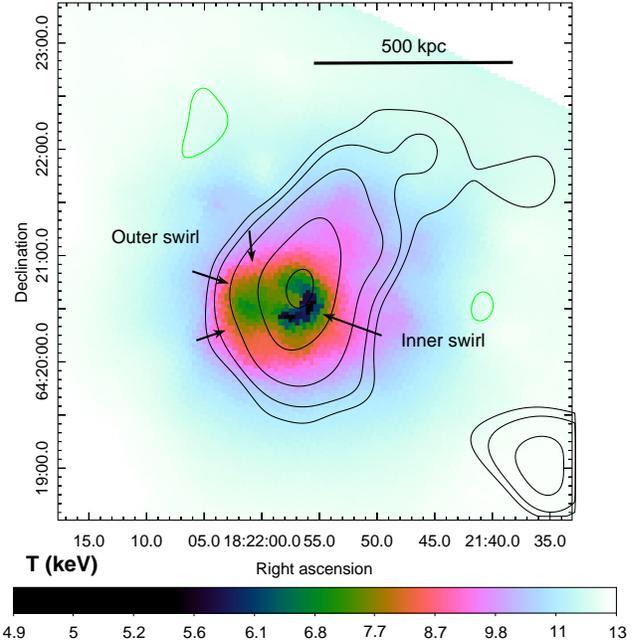}
 \caption{Temperature map shown with the radio halo contours (Fig.~\ref{h1821}, right) overlaid. 
 A colour version of the figure is available online.
\label{temp}
}
\end{figure}

The temperature map of CL1821+643 reveals two sub-structures which appear like `swirls' (Fig.~\ref{temp}). 
An inner swirl near the central quasar to the south-west and an outer 
swirl towards the east of the quasar at a distance of about 130 kpc is seen. 
The outer swirl is same as that reported by \citet{rus10}. Such swirls of cold gas 
have been seen in cool-core clusters and are proposed to be associated with the 
sloshing of low-entropy gas \citep{2014ApJ...795...73G}. The radio halo emission extends beyond the 
swirls but is less extended in the direction perpendicular (east-west) to the swirls  as compared to that 
parallel (north-south) to the swirls (Fig.~\ref{temp}).

\subsection{CL1821+643 and other clusters with radio haloes}
\begin{figure*}
\centering
\hspace*{-0.3in}
\includegraphics[width =\textwidth]{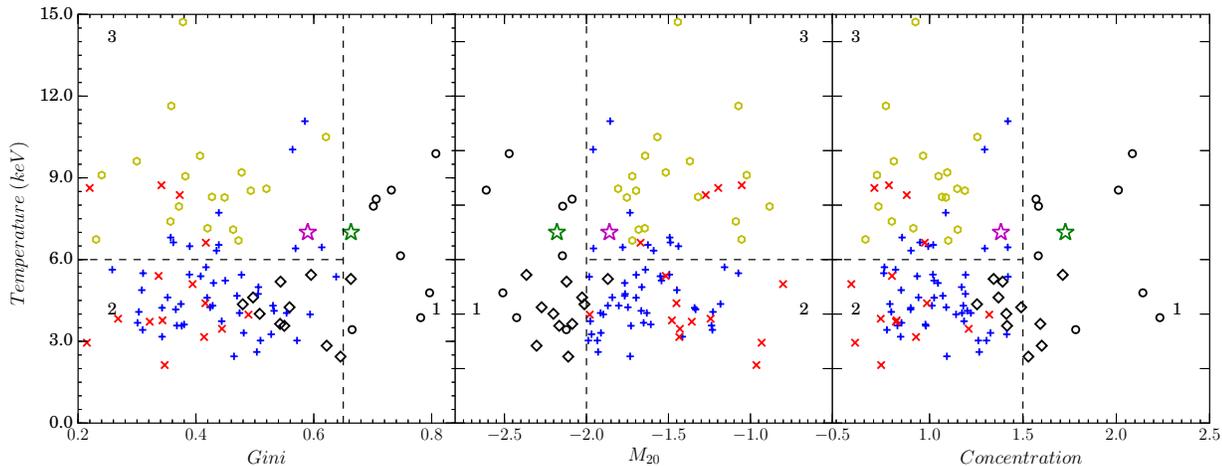}
\caption {\small Temperature versus morphological parameters plotted for P15 clusters and CL1821+643. 
Symbols are the same as Fig.\ref{CL1821+643_Xray1}. The regions marked 1, 2 and 3 
indicate the possible evolutionary sequence of a cluster from a strong relaxed cluster to 
a relaxed cluster and then to a strong non-relaxed cluster. The cluster CL1821+643 
without the quasar (magenta star) is located in the region of clusters with radio haloes.}
\label{tempmorf}
\end{figure*}

We also compared X-ray morphology of CL1821+643 with that of other clusters 
with known radio haloes. The morphological parameters (P15) for the 
sample of radio halo clusters from \citep{gio09} 
with $z>0.1$ is shown in Fig.~\ref{CL1821+643_Xray1} (hexagons). 
The clusters with radio halo are mainly in the NR and SNR categories and 
CL1821+643 fall in the NR category. It is more relaxed than other 
radio halo clusters.
We analyse the average cluster temperatures of the P15 cluster sample and CL1821+643 
in the context of presence of radio haloes. The average temperature of CL1821+643 is 
7 kev (B14). CL1821+643 along with P15 and the radio halo sample clusters
 are plotted in a temperature - morphological parameter plane (Fig.~\ref{tempmorf}).  
The plane is divided into regions 1, 2 and 3 in which the clusters seggregate.
 The three regions can also be viewed as an evolutionary sequence of clusters. 
The SR and R clusters are in region 1 and NR clusters are in region 2 and NR and SNR 
clusters are in region 3. The region 3 also contains most of the radio halo clusters.
This high temperature clusters showing highly disturbed ICM tend to 
host radio haloes. The cluster CL1821+643 (without quasar) is at the edge of the 
region 3, implying a disturbed ICM like that of other clusters with radio haloes.

\subsection{CL1821+643 and other clusters with mini-haloes}
\begin{figure}
\includegraphics[width=\columnwidth,angle=-90]{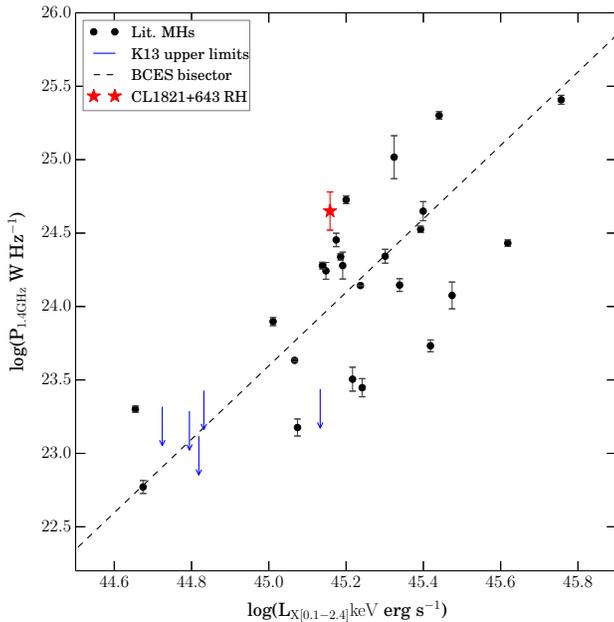}
 \caption{The CL1821+643 radio halo is shown with known mini-haloes in the \lxlr plane. 
 See \citet{kal13,2015A&A...579A..92K} and references therein for { the literature mini-haloes and the upper limits. 
 The best fit line $log(P_{\mathrm{1.4 GHz}}) = A \times log(\mathrm{L}_{X}) + B$ with $A=2.5\pm0.3$ and $B=-88.9\pm13.6$ is 
 shown.} 
 \label{mhlxlr}
}
\end{figure}

The radio halo in CL1821+643 is under-luminous in radio by a factor $\sim 3$ as 
compared to the expectation based on the correlation in the 
\lxlr plane for the radio haloes (B14). 
Since CL1821+643 also shows a cool-core near the quasar, we compare the radio halo
 with the known mini-haloes. {The 1.4 GHz power of the radio halo in CL1821+643, 
 $\mathrm{log}_{10}(P_{1.4\mathrm{GHz}} \mathrm{W Hz}^{-1}) = 24.65\pm0.05$, was calculated 
 assuming a spectral index of $1.0\pm0.1$ between 610 and 1400 MHz.}
 { The radio halo in CL1821+643 when plotted with other known mini-haloes in the \lxlr plane appears 
to be consistent (Fig.~\ref{mhlxlr}). 
 The surface brightness of the CL1821+643 is $1.7 \mu$Jy arcsec$^{-2}$. 
Both radio halos and mini-halos with similar surface brightness are known, for example, A1835 and 
A2254 \citep{mur09} and it falls in the region of overlap between the haloes and mini-haloes in surface brightness.
}

This radio halo is unlike other mini-halos that have been found confined to region between 
`cold-fronts \citep[e. g.][]{maz08, 2014ApJ...795...73G}.
According to recent MHD simulations 
the magnetic field structure in the cold-fronts 
can contain the mini-halo sizes to within the cold-front boundaries leading to 
a sharp fall in the surface brightness in the direction transverse to the 
cold-fronts; however allows further expansion 
in a direction parallel to the cold fronts \citep{zuh13, 2015ApJ...801..146Z}. 
In case of CL1821+643 radio 
halo is about 450 kpc in width in the direction transverse to the 
outer cold `swirl' (Fig.~\ref{temp}) but is nearly two times more in extent
in the northwest (see Sec.~\ref{rhmorph}). The swirls in temperature are similar to cold-fronts\citep{rus10} 
and the magnetic field in them can prevent further expansion in the 
transverse direction. The assymetry in the extent of the radio halo 
may be attributed to the presence of the cold-front-like swirls.

\section{Conclusions}\label{conclusion}
\begin{itemize}
\item { We have confirmed the radio halo in the cluster CL1821+643 at 610 MHz. Its
 flux density at 610 MHz is $35.6\pm4.0$ mJy and has an extent of $\sim 890$ kpc $\times 450$ kpc. 
  Equipartition magnetic field estimates of $0.63 \mu$G and $1.3 \mu$G were obtained under 
 standard assumptions following the methods using low frequency cut-off and low energy ($\gamma_{min}$) 
 cut-off, respectively}.
\item The integrated spectral index of the radio halo is  { $1.0\pm0.1$} over the frequency range 
is { 323 - 1665 MHz}.
\item The $Chandra$ archival data were 
reduced to make X-ray surface brightness and temperature { maps} of CL1821+643. 
The morphological parameters Concentration ($C$), Gini and $M_{20}$ were 
calculated using the surface brightness { map} and compared with those of a 
large sample of clusters in P15. CL1821+643 falls in the category of relaxed clusters when 
the central quasar is included and { in that of} non-relaxed clusters when the central region 
was excluded from the morphological analysis.
\item { CL1821+643 (quasar subtracted) is similar to other clusters with radio haloes in the temperature - 
morphological parameter plane.}
\item Being a cool-core cluster with a central quasar, we compared the radio halo with 
mini-halo clusters.{ The radio halo in the \lxlr plane for mini-haloes is 
consistent with other mini-haloes in radio power.}
\item { The larger extent of the radio halo in the north-south direction as compared to that in the 
east-west may be a consequence of the cosmic ray confinement by the magnetic fields in the 
cold swirls in the ICM of the cluster as proposed in recent simulations.}
\end{itemize}

\section*{Acknowledgments}
We thank the referee for critical comments that improved this paper. 
 RK is supported through the INSPIRE Faculty Award of the Department of Science and Technology (DST), India. 
We thank the staff of the GMRT, who have made these observations possible. 
GMRT is run by the National Centre for Radio Astrophysics of the Tata Institute
of Fundamental Research. 
The scientific results reported in this article are based 
 in part on 
 data obtained from the Chandra Data Archive 
  and published previously in cited articles.
This research made use of the NASA/IPAC
Extragalactic Database (NED), which is operated by the Jet Propulsion Laboratory,
California Institute of Technology, under contract with the National Aeronautics
and Space Administration. This research made use of data obtained from the High Energy Astrophysics
Science Archive Research Center (HEASARC), provided by NASA's Goddard Space
Flight Center.

\bsp

\label{lastpage}


\bibliographystyle{mn2e_new}
\bibliography{mybib}

\end{document}